\documentstyle[prl,aps,preprint]{revtex}
\begin{document}
\preprint{Phys. Rev. Lett. {\bf 72}, 713 (1994)} 
\draft 
\title{A Relation between the Correlation 
  Dimensions of Multifractal Wavefunctions and Spectral Measures in
  Integer Quantum Hall Systems} 
\author{Bodo Huckestein}
\address{Dept.\ of Electrical Engineering, Princeton University,
  Princeton, NJ 08540, USA} 
\author{Ludwig Schweitzer}
\address{Physikalisch-Technische Bundesanstalt, Bundesallee 100, 38116
  Braunschweig, Germany} 
\date{28 July 1993} 
\maketitle
\begin{abstract}
  We study the time evolution of wavepackets of non-interacting
  electrons in a two-dimensional disordered system in strong magnetic
  field. For wavepackets built from states near the metal-insulator
  transition in the center of the lowest Landau band we find that the
  return probability to the origin $p(t)$ decays algebraically, $p(t)
  \sim t^{-D_2/2}$, with a non-conventional exponent $D_2/2$.  $D_2$
  is the generalized dimension describing the scaling of the second
  moment of the wavefunction. We show that the corresponding spectral
  measure is multifractal and that the exponent $D_2/2$ equals the
  generalized dimension $\widetilde{D}_2$ of the spectral measure.
\end{abstract}

\narrowtext

Wavefunctions of critical states at a mobility edge separating
extended from localized states can be analyzed as multifractals
\cite{CP86,SP87,Eva90,GS91}. This also holds for Quantum Hall systems,
where the energy range of the extended states shrinks to a singular
point in the center of the Landau bands \cite{HK90,Huc92,HB92}. These
critical states exhibit a universal multifractal behavior that can be
described by an infinite set of generalized dimensions $D_q$ or
equivalently by a singularity strength distribution $f(\alpha)$
\cite{PJ91,HS92,HS92a}.  As a consequence of the multifractality of
the wavefunctions the two-particle spectral function $S(k,\omega)$ in
the quantum Hall system shows non-conventional
behavior\cite{Cha87,CD88}. At large values of $k^2/\omega$ the
diffusion constant as a function of wavevector $k$ and frequency
$\omega$ is reduced from its conventional value $D_0$, $D(k^2/\omega)
\sim D_0\,(k^2/\omega)^{-\eta/2}$, with $\eta=0.38\pm0.04$
\cite{CD88}. The exponent $\eta$ is related to the generalized
dimension $D_2$ of the wavefunction by $\eta=2-D_2$ \cite{PJ91,Cha90}. The
reduction in the diffusion constant will influence the long-time
behavior of autocorrelations of wavepackets built from states close to
the critical energy.

Another class of systems with multifractal wavefunction are
quasiperiodic systems like the self-dual Harper's equation
\cite{AA80,HiKo92}. In contrast to the Anderson transition where the
density of states is smooth these systems have multifractal spectra.
For quantum systems with Cantor spectra it was found that the temporal
autocorrelation function $C(t)$ of wavepackets built from the
multifractal eigenstates exhibits a slow algebraic decay, $C(t) \sim
t^{-\delta}$ \cite{KPG92}. Ketzmerick {\em et al.\/} showed that
$\delta$ equals the generalized dimension $\widetilde{D}_2$ of the
spectral measure introduced by the local density of states that form
the weights in the wavepackets.

In this paper we establish a connection between the multifractal
properties of the wavefunctions and the spectral measure for
wavepackets built from states near the critical energy in the center
of the lowest Landau band, namely that the generalized dimensions
$D_2$ of the wavefunction and $\widetilde{D}_2$ of the spectrum are
related by $\widetilde{D}_2=D_2/2$. In order to obtain this result we
study numerically the time evolution of the wavepackets for finite
systems. We find that the temporal autocorrelation function of these
wavepackets shows a slow algebraic decay $C(t) \sim t^{-\delta}$, with
$\delta=0.81\pm 0.02$.  Conventional, diffusive behavior would result
in $\delta=d/2=1$, with $d=2$ the euclidean dimension of the space. We
then show analytically that the multifractal structure of the
wavefunctions leads to the novel form $\delta=1-\eta/2=D_2/2$. By
numerically calculating the generalized dimension $\widetilde{D}_2$ of
the spectral measure we show that even in the presence of disorder
where the global density of states becomes smooth, the spectral measure
introduced by the local density of states near the critical energy is
multifractal and that as for Cantor spectra $\delta=\widetilde{D}_2$.
This allows us to connect the generalized dimensions of the
wavefunction and the spectral measure with the equality
$\widetilde{D}_2=D_2/2$.

The model that we use for the quantum Hall system is a two-dimensional
tight-binding model on a square lattice with on-site disorder and
transfer to nearest neighbors only
\begin{equation}
  {\cal{H}}=\sum_{m}\epsilon_{m}c^{\dagger}_{m}c^{}_{m} + \sum_{m\ne
    n} V_{mn} c^{\dagger}_{m} c^{}_{n}.
\end{equation}
The effect of the magnetic field is incorporated by a Peierls
substitution \cite{Pei33} via the vector potential ${\bf A(r)}$ in
the hopping matrix elements
\begin{equation}
  V_{mn} = V \exp\left(-ie/\hbar \int_{{\bf r_{m}}}^{{\bf r_{n}}}
  d{\bf r \,A(r)} \right).
\end{equation}

We used a system size of 125 by 125 lattice sites with periodic
boundary conditions and a commensurate magnetic field of 1/5 flux
quanta per unit cell of the lattice. This corresponds to a ratio of
system size to magnetic length $L/l_{c}$ of about 140. The disorder
potentials $\{\epsilon_{m}\}$ were taken from a constant distribution
of independent random variables with $-W/2 \le \epsilon_{m} \le W/2$.
The strength of the disorder ($W=3\,V$) was weak enough so that the
tight-binding band split into well separated Landau subbands.

In order to observe the multifractal properties of the integer quantum
Hall transition the system has to be effectively at the critical point
in the center of the Landau band. Approaching the critical energy
$E_c$ in the center of the Landau band the localization length
$\xi(E)$ diverges. In a finite system the transition is rounded and
the system is effectively critical when the localization length
$\xi(E)$ of the states under consideration is large compared to the
system size $L$.  Over the energy interval used in our calculation the
localization length exceeds the system size by at least a factor of 2
\cite{HK90,Huc92}. Within this critical region no systematic energy
dependence of the generalized dimensions was observed.

We used 330 states (about 10\,\% of a Landau band) near the center of
the lowest Landau level to build the wavepackets. We construct
normalized wavepackets $\psi({\bf r},t)$ from the eigenstates
$\phi_i({\bf r})$ with energy $E_i$ so that at time $t=0$ they are
centered around ${\bf r}={\bf 0}$,

\begin{equation}
  \psi({\bf r},t) = A^{-1/2} \sum_i \phi_i^*({\bf 0})
                                  \phi_i({\bf r}) e^{i E_i t/\hbar},
  \label{packet}
\end{equation}
with $A=\sum_i |\phi_i({\bf 0})|^2$.  The probability density
$p(t)$ to find the particle at site ${\bf r}={\bf 0}$ at time
$t$ is then given by

\begin{equation}
  p(t) = |\psi({\bf 0},t)|^2 = A^{-1} \sum_{i,j}
              |\phi_i({\bf 0})|^2 |\phi_j({\bf 0})|^2
                   e^{i (E_i - E_j) t/\hbar}.
  \label{p(t)}
\end{equation}

This quantity can also be interpreted (up to a constant) as the
probability to find the wavepacket $\psi({\bf r},t) = \langle
{\bf r}| \psi(t) \rangle$ at time $t$ in the initial state
$|\psi(0)\rangle$, $p'(t) = A^{-1} p(t) = |\langle \psi(t=0)| \psi(t)
\rangle|^2$. A temporal autocorrelation function $C(t)$ was defined by
Ketzmerick {\em et al.\/} by smoothing $p'(t)$ \cite{KPG92}

\begin{equation}
  C(t) = \frac{1}{t}\int_0^t dt' p'(t').
  \label{C(t)}
\end{equation}

As can be seen from Fig.~(\ref{fig:C(t)}) $C(t)$ decays algebraically,
$C(t) \sim t^{-\delta}$, with $\delta = 0.81 \pm 0.02$. Conventional
diffusion would lead to a wavepacket with asymptotically gaussian
shape and a probability $p'(t) \sim t^{-d/2}$, where $d$ is the
euclidean dimension of the space. The initial deviations from the
power law decay seen in Fig.~(\ref{fig:C(t)}) are due to the intrinsic
width of the wavepacket at $t=0$, whereas the behavior at times
$t>8000$ arises from finite size effects.

The slow decay of the temporal autocorrelation function is a result of
the multifractal structure of the wavefunctions and is related to the
structure of the two-particle spectral function $S(k,\omega)$ \cite{CD88},

\begin{equation}
  S(k,\omega) = \rho(E_c)\frac{\hbar}{\pi}\frac{k^2
    D(k^2/\omega)}{\omega^2 + \hbar^2 k^4 (D(k^2/\omega))^2}.
  \label{sqo1}
\end{equation}
In the limit $k,\omega $ going to zero and $k^2/\omega$ large compared
to the density of states at the critical energy $\rho (E_{c})$,
$S(k,\omega)$ is proportional to $\omega^{-\eta/2}k^{-2+\eta}$
\cite{CD88}. The exponent $\eta$ is related to the generalized
dimension $D_2$ of the wavefunction by $\eta=2-D_2$ \cite{PJ91}.
Fourier transforming the spectral function and taking the limit
${\bf r}\to {\bf 0}$ gives the probability density $p(t)$,

\begin{equation}
  p(t) =\frac{1}{\rho(E_c)} \int_{-\infty}^{\infty}
  d\omega\,e^{i\omega t} \int_{1/L}^{1/l_{m}} dk\, k S(k,\omega),
  \label{fourier}
\end{equation}
where the momentum integral is cut off at small wavelengths by the
microscopic length $l_{m}$ (i.e.\ magnetic length or lattice spacing)
and at long wavelengths by the system size $L$.  The long time limit
of Eq.~(\ref{fourier}) is governed by the small frequency behavior of
the momentum integral which in turn is determined by the large
$k^2/\omega$ limit of $S(k,\omega)$,

\begin{eqnarray}
  \int_{1/L}^{1/l_{m}} dk\, k S(k,\omega) & = & \int_{1/\omega
    L^2}^{1/\omega l^2_{m}} d(k^2/\omega)\,\frac{\omega}{2}
  S(k,\omega)\nonumber\\ & \sim & \int_{1/\omega L^2}^{1/\omega
    l^2_{m}} d(k^2/\omega)\,\frac{\omega}{2}
  \omega^{-\eta/2}k^{-2+\eta} \nonumber\\ & \sim & \omega^{-\eta/2},
  \label{fou2}
\end{eqnarray}
so that in the limit $t \to \infty$ the probability density $p(t)$
becomes

\begin{equation}
  p(t) \sim \int_{-\infty}^{\infty} d\omega e^{i\omega
    t}|\omega|^{-\eta/2} \sim \frac{1}{t^{1-\eta/2}} =
  \frac{1}{t^{D_2/2}}.
  \label{ptresult}
\end{equation}

Thus the temporal autocorrelations decay as if the wavefunctions would
show conventional diffusive behavior but in a fractal
$D_2$-dimensional space instead of the euclidean two-dimensional
space. This interpretation is further supported by observing that the
slow decay of the probability density $p(t)$ is solely due to the
shape of the wavepacket becoming non-gaussian. The variance of the
wavepacket still grows proportional to $t$ as is the case for
solutions to the diffusion equation in arbitrary dimension.
Fig.~(\ref{variance}) shows the variance $R(2,t) = \int d^2r |{\bf
  r}|^2 |\psi({\bf r},t)|^2$ of the wavepacket as a function of
time $t$.

The local density of states $|\phi_i({\bf 0})|^2$ used as the
weights of the eigenfunctions $\phi_i({\bf r})$ in constructing
the wavepacket introduces a spectral measure. Due to the eigenfunction
correlations this measure is multifractal even though the global
density of states is non-critical and smooth near the metal-insulator
transition.

While the present calculation explains the slow decay of temporal
autocorrelations in terms of the multifractal properties of the
wavefunction, Ketzmerick {\em et al.\/} have related this phenomenon in
quantum systems with Cantor spectra to the multifractal properties of
the spectral measure \cite{KPG92}. They showed that $\delta$ is given by
the generalized dimension $\widetilde{D}_2$ of the spectral measure.
We calculated $\widetilde{D}_2$ for the same energy interval that we
used for the wavepackets from \cite{KPG92}

\begin{equation}
  \gamma(\varepsilon) = \sum_{i} \left( A^{-1} \sum_{E \in \Omega
    _{i}(\varepsilon)} |\phi_{E}({\bf 0})|^{2}\right)^{2} \sim
  \varepsilon^{\widetilde{D}_{2}},\ (\varepsilon \to 0),
  \label{dtwiddle}
\end{equation}
where the energy interval is partitioned into boxes
$\Omega_{i}(\varepsilon)$ of width $\varepsilon $.

Fig.~(\ref{spectrum}) shows that even in the presence of disorder the
spectral measure at the center of the lowest Landau band is
multifractal with a generalized dimension $\widetilde{D}_2=0.8\pm
0.05$. Thus the relation $\delta=\widetilde{D}_2$ holds for the
Quantum Hall system, too. This allows us to directly relate the
generalized dimensions $D_2$ of the wavefunction and $\widetilde{D}_2$
of the spectral measure, $\widetilde{D}_2=D_2/2$.  This is to our
knowledge the first time that such a direct connection could be made.

The correlation dimension $D_{2}$ was obtained \cite{HP83} from the
scaling of $P_{2}(\lambda) \sim \lambda^{D_{2}} $, see Fig.~(\ref{d2}),
where $l=\lambda L$ is the length of the boxes $\Omega_i(\lambda)$
used to cover the fractal eigenstate with energy $E$ and

\begin{equation}
  P_{2}(\lambda ,E)=\sum_{i}\left(\sum_{{\bf r} \in
    \Omega_{i}(\lambda)}|\phi _{E}({\bf r})|^{2}\right)^{2}.
  \label{pq}
\end{equation}

In conclusion, we have established a relation between the multifractal
properties of wavepackets built from critical eigenstates near the
center of the lowest Landau level of a Quantum Hall system and the
multifractal properties of the corresponding spectral measure.
Specifically we find that the generalized dimensions $D_2$ of the
wavefunction and $\widetilde{D}_2$ of the spectral measure are related
by $\widetilde{D}_2=D_2/2$. As a consequence of these multifractal
properties the temporal autocorrelation function $p(t)$ decays with a
non-conventional exponent $\delta=D_2/2$. Thus the probability for the
wavepacket staying near the origin after long times is greatly
enhanced compared to conventional diffusion in two dimension. This
behavior can be interpreted as a wavepacket showing conventional
diffusion on a $D_2$-dimensional fractal.

One of us (B.H.) gratefully acknowledges the support of a NATO
scholarship provided through the DAAD and a DFG scholarship and the
hospitality of the Aspen Center for Physics during the final stages of
this work.

%\bibliographystyle{prsty}
%\bibliography{bodos,ludwig,fracbod}

\begin{thebibliography}{10}

\bibitem{CP86} C. Castellani and L. Peliti, J. Phys. A: Gen. {\bf 19},
L429 (1986).

\bibitem{SP87}
A.~P. Siebesma and L. Pietronero, Europhys. Lett. {\bf 4},  597  (1987).

\bibitem{Eva90}
S.~N. Evangelou, J. Phys. A: Math. Gen. {\bf 23},  L317  (1990).

\bibitem{GS91}
M. Schreiber and H. Grussbach, Phys.\ Rev.\ Lett. {\bf 67},  607  (1991).

\bibitem{HK90}
B. Huckestein and B. Kramer, Phys.\ Rev.\ Lett. {\bf 64},  1437  (1990).

\bibitem{Huc92}
B. Huckestein, Europhys. Lett. {\bf 20},  451 (1992).

\bibitem{HB92}
Y. Huo and R. Bhatt, Phys.\ Rev.\ Lett. {\bf 68},  1375  (1992).

\bibitem{PJ91}
W. Pook and M. Jan\ss{}en, Z. Phys.\ B {\bf 82},  295  (1991).

\bibitem{HS92}
B. Huckestein and L. Schweitzer, Helvetica Physica Acta {\bf 65},  317  (1992).

\bibitem{HS92a}
B. Huckestein and L. Schweitzer,  in {\em High Magnetic Fields in Semiconductor
  Physics III}, Vol.~101 of {\em Springer Series in Solid State Sciences},
  edited by G. Landwehr (Springer-Verlag, Berlin, 1992), pp.\ 84--88.

\bibitem{Cha87}
J.~T. Chalker, J. Phys.\ C {\bf 20},  L493  (1987).

\bibitem{CD88}
J.~T. Chalker and G.~J. Daniell, Phys.\ Rev.\ Lett. {\bf 61},  593  (1988).

\bibitem{Cha90}
J.~T. Chalker, Physica A {\bf 167}, 253 (1990).

\bibitem{AA80}
S. Aubry and G. Andre, Ann.\ Isr.\ Phys.\ Soc. {\bf 3},  133  (1980).

\bibitem{HiKo92}
H. Hiramoto and M. Kohmoto, Int. J. Mod. Phys. B {\bf 6},  281  (1992).

\bibitem{KPG92}
R. Ketzmerick, G. Petschel, and T. Geisel, Phys.\ Rev.\ Lett. {\bf 69},  695
  (1992).

\bibitem{Pei33}
R. Peierls, Z. Physik {\bf 80},  763  (1933).

\bibitem{HP83}
H.~G.~E. Hentschel and I. Procaccia, Physica {\bf 8D},  435  (1983).

\end{thebibliography}

\begin{figure}
\caption{\label{fig:C(t)} Temporal autocorrelation function $C(t)$
  vs.\ time $t$ (in units of $\hbar/V$) showing the non-conventional
  power law behavior $C(t) \sim t^{-\delta}$ with
  $\delta=0.81\pm0.02$. For comparison the dashed line shows the
  conventional behavior $C(t) \sim t^{-1}$.}
\end{figure}

\begin{figure}
%input rv2tm125.pic
  \caption{\label{variance} The variance $R(2,t)$ of the wavepacket as
    a function of time $t$ showing conventional diffusive growth
    proportional to $t^{\kappa}$ with an exponent $\kappa =1.0\pm0.04$.}
\end{figure}

\begin{figure}
\caption{\label{spectrum} The scaling of the second moment
  $\gamma(\varepsilon)$ of the spectral measure in the vicinity of the
  transition (Eq.~(\protect\ref{dtwiddle})) with fractal exponent
  $\widetilde{D}_{2}=0.8\pm0.05$ that coincides with the exponent
  $\delta$ of Fig.~(\protect\ref{fig:C(t)}) within the statistical
  errors.}
\end{figure}

\begin{figure}
\caption{\label{d2} The scaling of the second moment $P_2(\lambda)\sim
  \lambda^{D_{2}}$ of critical states near the transition with the
  correlation dimension $D_{2}=1.62\pm 0.02$ fulfilling the relation
  $\widetilde{D}_2=D_2/2$ within the statistical errors.} \end{figure}
\end{document}